\documentclass{jfm}

\usepackage{graphicx}
\usepackage{natbib}

\ifCUPmtlplainloaded \else
  \checkfont{eurm10}
  \iffontfound
    \IfFileExists{upmath.sty}
      {\typeout{^^JFound AMS Euler Roman fonts on the system,
                   using the 'upmath' package.^^J}%
       \usepackage{upmath}}
      {\typeout{^^JFound AMS Euler Roman fonts on the system, but you
                   dont seem to have the}%
       \typeout{'upmath' package installed. JFM.cls can take advantage
                 of these fonts,^^Jif you use 'upmath' package.^^J}%
      }
  \else
  \fi
\fi

\ifCUPmtlplainloaded \else
  \checkfont{msam10}
  \iffontfound
    \IfFileExists{amssymb.sty}
      {\typeout{^^JFound AMS Symbol fonts on the system, using the
                'amssymb' package.^^J}%
       \usepackage{amssymb}%
         \let\leq=\leqslant
       \let\ge=\geqslant  
      }{}
  \fi
\fi

\ifCUPmtlplainloaded \else
  \IfFileExists{amsbsy.sty}
    {\typeout{^^JFound the 'amsbsy' package on the system, using it.^^J}%
     \usepackage{amsbsy}}
    {\providecommand\boldsymbol[1]{\mbox{\boldmath $##1$}}}
\fi

\providecommand\bnabla{\boldsymbol{\nabla}}

\title[Dimensional transition in stratified thin fluid layers]
{Dimensional transition of energy cascades in stably stratified thin 
fluid layers}

\author[A. Sozza, G. Boffetta, P. Muratore-Ginanneschi, S. Musacchio]
{A. Sozza$^1$,
G. Boffetta$^1$,
P. Muratore-Ginanneschi$^2$,
S. Musacchio$^3$}

\affiliation{$^1$Department of Physics and INFN, University of Torino,
via P.Giuria 1, 10125 Torino, Italy \\[\affilskip]
$^2$Department of Mathematics and Statistics, University of Helsinki PL 68, FIN-00014 Helsinki, Finland
\\
$^3$Universit\'e de Nice Sophia Antipolis, CNRS,  
LJAD, UMR 7351, 06100 Nice, France}

\begin{document}
\maketitle

\begin{abstract}
Numerical simulations of a thin layer of turbulent flow in stably
stratified conditions within the Boussinesq approximation have
been performed. The statistics of energy transfer among scales have
been investigated for different values of control parameters: 
thickness of the layer and density stratification. 
It is shown that in a thin layer with a quasi-two-dimensional 
phenomenology, stratification provides a new channel for the energy 
transfer towards small scales and reduces the inverse cascade.
The role of vortex stretching and enstrophy flux in the transfer
of kinetic energy into potential energy at small scales is discussed.
\end{abstract}


\section{Introduction}
\label{sec1}

In many instances of geophysical flows, 
the fluid motion is confined by material boundaries 
or other physical mechanisms in thin layers with a small aspect ratio.  
The thickness of such layers can be much smaller than 
the typical horizontal scales of motion, 
while being at the same time much larger than the dissipative viscous scales.

The turbulent dynamics of flows confined in quasi two-dimensional geometries 
exhibits a rich and interesting phenomenology. 
Numerical simulations \citep{SCW96, CMV10} 
and experiments \citep{SBX10,XBFS11} have shown that a mixture 
of two-dimensional (2D) and three-dimensional (3D) 
dynamics can emerge in such situation.  
In particular, when the thickness of the fluid layer is smaller 
than the length scale of the forcing 
the turbulent cascade of kinetic energy injected 
by external forces splits in two parts. 
A fraction of the energy is transferred toward small, 
viscous scales as in 3D turbulence, 
while the remnant energy undergoes an inverse cascade 
toward large scales as in 2D turbulence. 
The key parameter which determines the energy flux 
of the two cascades is the ratio $S=L_z/L_f$
between the confining scale $L_z$ 
and the length scale of the forcing $L_f$ 
\citep{SCW96, CMV10}. 

Beside the confinement, other physical factors can
affect the effective dimensionality of geophysical flows.
One important ingredient is rotation, which 
typically favors a two-dimensionalization
of the flow \citep{SW99,PSRMB13,DBLM14}.
The presence of a stable stratification of density 
also affects the dimensionality of geophysical flows 
(for a review see, e.g.~\cite{RL00}). 
A typical feature of strongly stratified turbulent flows, 
which has been observed both in numerical simulations and experiments 
(\cite{BC00,BC01,SW02,WB04,GCB04,PFS05,BBLC07} among others),
is the formation of quasi-horizontal layered structure, 
often called ``pancakes''. 
It has been argued by \cite{BC00} that their formation could 
be connected to instabilities of columnar vortices.

The presence of these structures 
is accompanied by a breaking of the isotropy of the flow. 
In particular the vertical velocities are strongly suppressed
and the horizontal flows become predominant. 
At the same time the strong shear between 
the layered structures originates intense vertical 
gradients of horizontal velocities.  
Scaling analysis by \cite{BC01} and \cite{L06} predicts that 
the typical thickness of the layered structures $L_v$ 
is proportional to the magnitude of horizontal velocity $U$ 
and to the inverse of the Brunt-V\"ais\"ala frequency $N$. 
The aspect ratio between the vertical scale $L_v$ 
and horizontal scale $L_h$ of the flow is therefore proportional 
to the horizontal Froude number $L_v/L_h \sim F_h = U/NL_h$. 

The role played by these structures on the transfer of energy 
in stratified turbulence has been extensively investigated because 
of its relevance for the dynamics of the atmosphere. 
Early studies by \cite{L83} proposed the hypothesis that 
two-dimensional turbulence could develop within the horizontal layers, 
leading to the formation of an inverse energy cascade 
which could transfer kinetic energy toward large-scale structures. 
On the other hand, predictions based on 
eddy damped quasi-normal Markovian (EDQNM) closure 
by \cite{GC94} have argued that 
the formation of layered structures should block the development 
of an inverse energy cascade because of the strong dissipation 
due to the turbulent shear between layers. 

Although early simulations of stratified flows
by \cite{HM89} reported the presence of a weak inverse cascade, 
most numerical simulations (\cite{WB04,WB06,L06,BBLC07,MMRP13})
have found a direct cascade of kinetic energy.
An energy transfer from small to large scale has been observed 
in numerical simulations by \cite{SW02,LMD03,WB04,WB06,L06} 
for Froude number smaller than a $O(1)$ critical value, 
but this phenomenon is not associated to an  
inverse energy cascade of vortical energy as in 2D turbulence. 
Conversely, kinetic energy piles up in vertically sheared horizontal 
flows (VSHF), 
i.e., shear modes with $k_h=0,k_z \neq 0$. 
It has been argued that this process is the result of 
inertial-gravity waves interactions, 
but its mechanisms are still poorly understood.  
The situation is even more complex when the stratified flow 
is subject to rotation. In stably stratified rotating flows, 
an inverse cascade has been observed \citep{MBGRL96,SW02,WB06,MMRP13}. 
In particular, when rotation and stratification 
have comparable strength (i.e. rotation rate and Brunt-Vaisala frequency
are similar), the energy transfer toward large scales 
is found to be larger than in the purely rotating case \citep{MMRP13}.  

The picture which emerge from the 
results of previous studies is that in the limit of strong stratification, 
and in absence of rotation,  
the turbulent flow is characterized by a cascade toward small scales both 
of kinetic energy and potential energy (see, e.g., \cite{L06,BBLC07}).
These observations pose an intriguing question. 
Given the fact that in absence of stratification the turbulent dynamics 
of thin fluid layers shows the development of an inverse cascade, 
which on the contrary is absent in presence of a strong stratification, 
it is natural to investigate the physical mechanism which
bridges these two extremes cases. 

In this work we consider the consequences of stable stratification on 
a thin layer of turbulent flow and the effect on the direction of the 
energy cascade.
By means of a set of high-resolution numerical simulations, we
investigate the mechanisms of transfer of kinetic and potential energy in a
stably stratified turbulent flow forced at intermediate scales and 
confined in fluid layers with variable aspect ratio. 
We show that the turbulent cascade of potential energy acquired by the fluid, 
acts as a new channel for dissipation 
which reduces the large-scale flux of kinetic energy 
and eventually prevents the development of the inverse cascade.

The remaining of the paper is organized as follows: 
Section~\ref{sec2} presents the governing equation and
described the numerical simulations.
In Section \ref{sec3} we report and discuss the results obtained. 
Section \ref{sec4} is devoted to conclusions.  

\section{Governing equations and numerical simulations}
\label{sec2}
The equation of motion for an incompressible flow, 
stably stratified in the vertical direction 
by a mean density gradient $\gamma$, 
in the gravitational field ${\bf g}=(0,0,-g)$
within the Boussinesq approximation are
\begin{eqnarray}
{\partial {\bf u} \over \partial t} + {\bf u} \cdot \bnabla {\bf u} &=&
- {1 \over \rho_0} \bnabla p + \nu \nabla^2 {\bf u} - N \phi {\bf e}_3 
+ {\bf f}
\label{eq:2.1} \\
{\partial \phi \over \partial t} + {\bf u} \cdot \bnabla \phi &=&
\kappa \nabla^2 \phi + N {\bf e}_3 \cdot {\bf u} 
\label{eq:2.2} 
\end{eqnarray}
supplemented by the condition of incompressibility $\bnabla \cdot {\bf u}=0$. 
Here, ${\bf u}=(u_1,u_2,u_3)$ is the 
velocity field, 
$\phi$, which has the dimension of a velocity,
is proportional to the deviation of the density field 
$\rho$ from the linear vertical ($z$ direction) profile 
$\rho=\rho_0+\gamma (\phi/N-z)$, 
$\gamma$ is the mean density gradient
which, together with the acceleration of gravity $g$, defines 
the Brunt-Vaisala frequency $N$ as $N^2=\gamma g/\rho_0$.
The parameters $\nu$ and $\kappa$ are the molecular viscosity 
and diffusivity respectively.
The flow is sustained by the external force ${\bf f}({\bf x},t)$, 
which is active on a characteristic correlation scale $L_f$ and 
provides an energy input rate $\varepsilon_f$.  

In the absence of forcing and dissipations, equations 
(\ref{eq:2.1}-\ref{eq:2.2}) conserve the total energy density,
given by the sum of kinetic and potential contributions: 
\begin{equation}
E=E_k + E_p = {1 \over 2} \langle |{\bf u}|^2 \rangle +
{1 \over 2} \langle \phi^2 \rangle
\label{eq:2.4}
\end{equation}
where $\langle ... \rangle={1 \over V} \int d^3 x ...$ 
represents the average over the whole volume $V$ of the fluid.  

The energy balance (see Appendix~\ref{appA} for its derivation)
reads: 
\begin{equation}
{dE_{K} \over dt}=\varepsilon_{f}-\varepsilon_{\nu}-\varepsilon_{x}~,
\qquad\quad
{dE_{P} \over dt}=\varepsilon_{x}-\varepsilon_{\kappa}
\end{equation}
where we have introduced the viscous dissipation rate of kinetic energy 
$\varepsilon_\nu = \nu \langle \partial_i u_j \partial_i u_j \rangle $
(summation over repeated indices is assumed),
the diffusive dissipation rate of potential energy 
$\varepsilon_{\kappa} = \kappa \langle \partial_i \phi \partial_i \phi \rangle$
and the exchange rate between kinetic and potential energy
$\varepsilon_{x}=N\langle{u_3 \phi}\rangle$. 
The sign of the exchange rate $\varepsilon_x$ is not defined a priori, 
but the analysis of the K\'arm\'an-Howart-Monin equations 
(discussed in Appendix~\ref{appA}) shows that it is positive on average. 
This indicates that it causes an effective dissipation of kinetic energy 
and a production of potential energy, which is indeed observed in our 
simulations.

In the limit of vanishing stratification, $N \to 0$ 
the field $\phi$ decouples from the velocity field in (\ref{eq:2.1}) 
which therefore recovers the usual Navier-Stokes equation, 
while (\ref{eq:2.2}) becomes the equation for the evolution 
of a passive scalar field (which therefore does not affect the 
velocity field).
Because in this limit (\ref{eq:2.2}) is not forced, the scalar field
eventually vanishes and the energy (\ref{eq:2.4}) is dominated by
the kinetic component only, which becomes an inviscid invariant.

In this limit $N=0$, the transition from 2D to 3D turbulent
phenomenology is determined by the aspect ratio $S=L_z/L_f$ between the
vertical scale and the characteristic forcing scale $L_f$ 
\citep{SCW96,CMV10}.
As $S \to 0$ one recovers the 2D scenario in which the energy flux is
towards the large scales $\ell \gg L_f$ and 
kinetic energy grows linearly at a rate
$\varepsilon_{\alpha}=dE_{K}/dt$. 
In the limit of very small viscosity
the standard 2D phenomenology \citep{BE12} predicts that 
$\varepsilon_{\alpha}=\varepsilon_f$, i.e. all the injected energy is 
transferred to large scale.
Numerical simulations at increasing values of $S$ have shown \citep{CMV10}
a progressive transition toward the 3D phenomenology of direct cascade.
When the vertical scale $L_z$ becomes larger than the viscous scale, 
an intermediate regime with 
both a direct and an inverse cascade of kinetic energy appears. 
In this "split cascade" regime, intermediate between 2D and 3D turbulence, 
a fraction $\varepsilon_{\alpha}<\varepsilon_f$ of energy input is 
transferred toward large scales 
while the remaining energy is transferred toward small scales
where it is dissipated by viscosity at a rate $\varepsilon_{\nu}$. 
The conservation of kinetic energy gives a constraint 
for the sum of the fluxes of the inverse and direct energy cascade 
$\varepsilon_{\alpha}+\varepsilon_{\nu}=\varepsilon_f$.
By increasing the aspect ratio $S$ the ratio 
$\varepsilon_{\nu}/\varepsilon_{\alpha}$
increases and eventually a complete 3D phenomenology
is recovered in which the system attains a stationary state with 
$\varepsilon_{\alpha}=0$.
The value of $S$ at which the flux of the inverse cascade vanishes
is not universal and may depend on the detail of the forcing. 
In the numerical setup adopted by \cite{CMV10} this is observed
for $S \approx 1/2$. 

The main goal of our work is to investigate how the above scenario  
is affected by the presence of a stable stratification of density in 
the fluid layer. 
In particular we will study how the various quantities which appears 
in the energy balance depends both on the aspect ratio $S$ and the 
Froude number $Fr$, and we will determine which part of the 
parameter space $(S,Fr)$ allows for the development of an inverse 
energy cascade. 

\subsection{Numerical simulations}

We performed direct numerical simulations of equations 
(\ref{eq:2.1}-\ref{eq:2.2})
by means of a parallel, fully dealiased, pseudo-spectral code. 
The flow is confined in a triply-periodic domain, with horizontal sizes
$L_x=L_y=2\pi$ and vertical size $L_z \ll L_x, L_y$. 
The domain is discretized on a regular grid with resolution $N_x = N_y = 512$
and $N_z = N_x (L_z/L_x)$. 

In the setup of our numerical simulations, 
the forcing ${\bf f}({\bf x},t)$ which sustains the turbulent flow 
is a Gaussian, white-in-time 
stochastic noise, active only on the horizontal components of velocity 
$(u_1,u_2)$ and depending on the horizontal components $(x,y)$ only.
This choice is consistent with the two-dimensional limit for $L_z \to 0$
and gives a forcing (and energy input) which is independent on 
the parameter $L_z$.
The forcing is localized in the Fourier space in a narrow band of wave numbers
$|{\bf k}|\simeq{k_{f}}=8$ and injects energy into the system at a
fixed rate $\varepsilon_{f}$. 
We adopt a hyperviscous damping scheme 
$\nu_{p}\nabla^{2p}$ of order $p=8$ both for the viscosity and the diffusivity, 
with $\nu_p/\kappa_p =1$. 

The non-dimensional parameter which quantify the confinement 
is given by the ratio 
\begin{equation}
\label{eq:2.s}
S=\frac{L_z}{L_f} 
\end{equation}
between the vertical scale and the characteristic forcing scale 
$L_f=2\pi/k_f$. 

The intensity of the stratification is expressed in term of the 
Froude number, a second dimensionless number here defined as in \cite{SW02}:  
\begin{equation}
\label{eq:2.fr}
Fr=\frac{\varepsilon_{f}^{1/3}k_{f}^{2/3}}{N}
\end{equation}
It is worth to remind that this definition of the Froude number 
is aimed to non-dimensionalize the parameters which are fixed 
a priori in the simulations, 
and it is not based on observables which are measured a posteriori. 
The region of parameter space $(S,Fr)$ explored in our simulations 
is shown in Table~\ref{table1}. 

\begin{table}
\begin{tabular}{c||c|c|c|c|c|c|c|c|c|c|c}
$Fr \backslash S$ & $0.5$ & $0.438$ & $0.375$ & $0.344$ & $0.313$ & $0.282$ & $0.25$ & $0.219$ & $0.188$ & $0.172$ & $0.125$ \\ \hline
$\infty$  & $0.035 $  & $0.054$  & $0.099$  &         & $0.160$   &       & $0.282$   &         & $0.384$   &          & $0.540$   \\
$0.75$    &         & $0.032$  & $0.060$  &         & $0.186$   &         & $0.271$   &         & $0.341$   &         & $0.526$   \\
$0.5$     &         & $0.025$  & $0.024$  &         & $0.138$   &         & $0.251$   &         & $0.329$   &         & $0.527$   \\
$0.4$     &         &         & $0.038$  &          & $0.117$   &         & $0.234$   &         & $0.312$   &         & $0.459$   \\
$0.3$     &         &         &         & $0.0$     & $0.047$  & $0.089$  & $0.132$   & $0.246$   & $0.274$   &         & $0.478$   \\
$0.25$    &         &         &         &         & $0.0$     & $0.044$  & $0.112$   & $0.208$   & $0.214$   &         & $0.450$   \\
$0.2$     &         &         &         &         &         &         & $0.0$     & $0.098$   & $0.150$   & $0.275$   & $0.415$   \\
$0.15$    &         &         &         &         &         &         &         &         & $0.002$  & $0.123$   &        
\end{tabular}
\caption{Parameter space of the simulations. Each number 
corresponds to a simulation and 
represents the fraction of energy transferred 
to large scales, $\varepsilon_{\alpha}/\varepsilon_f$, as a function of
$Fr=(\varepsilon_{f}^{1/3}k_{f}^{2/3})/N$ and $S=L_z/L_f$.}
\label{table1}
\end{table}

\section{Results}
\label{sec3}

\begin{figure}
\begin{center}
\includegraphics[width=8cm]{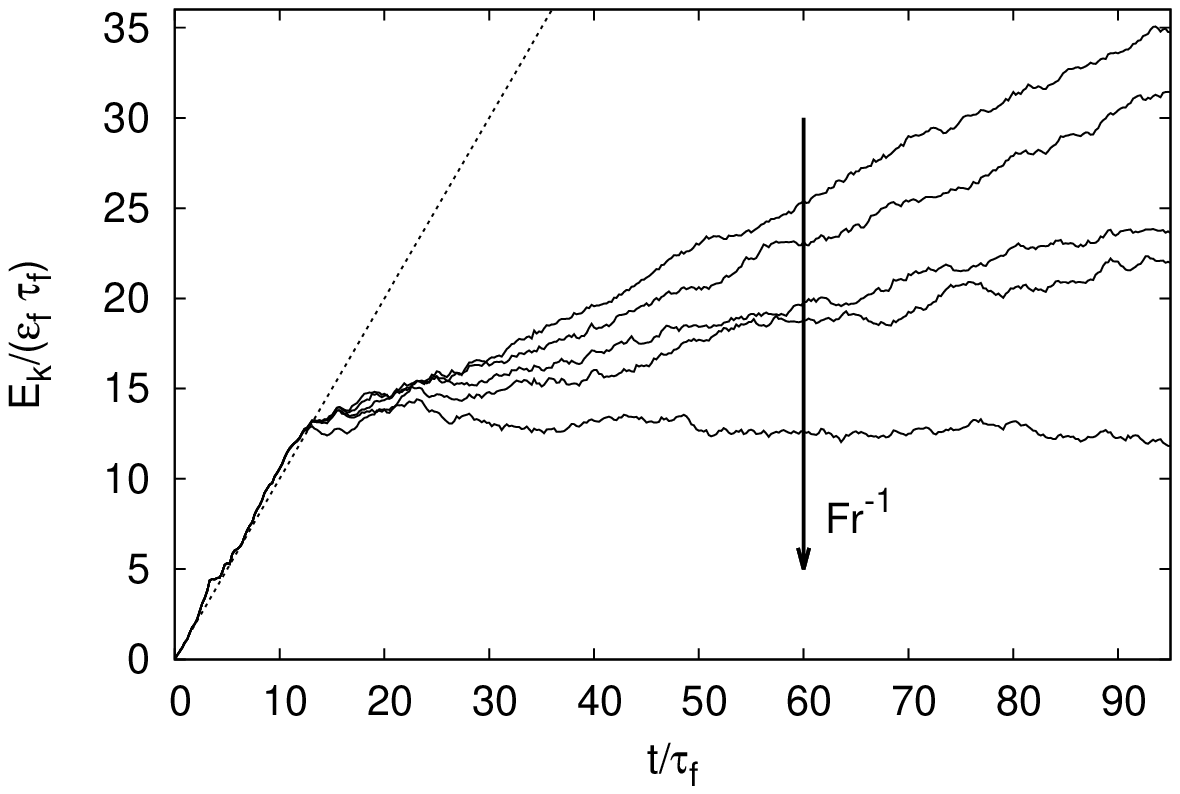}
\includegraphics[width=8cm]{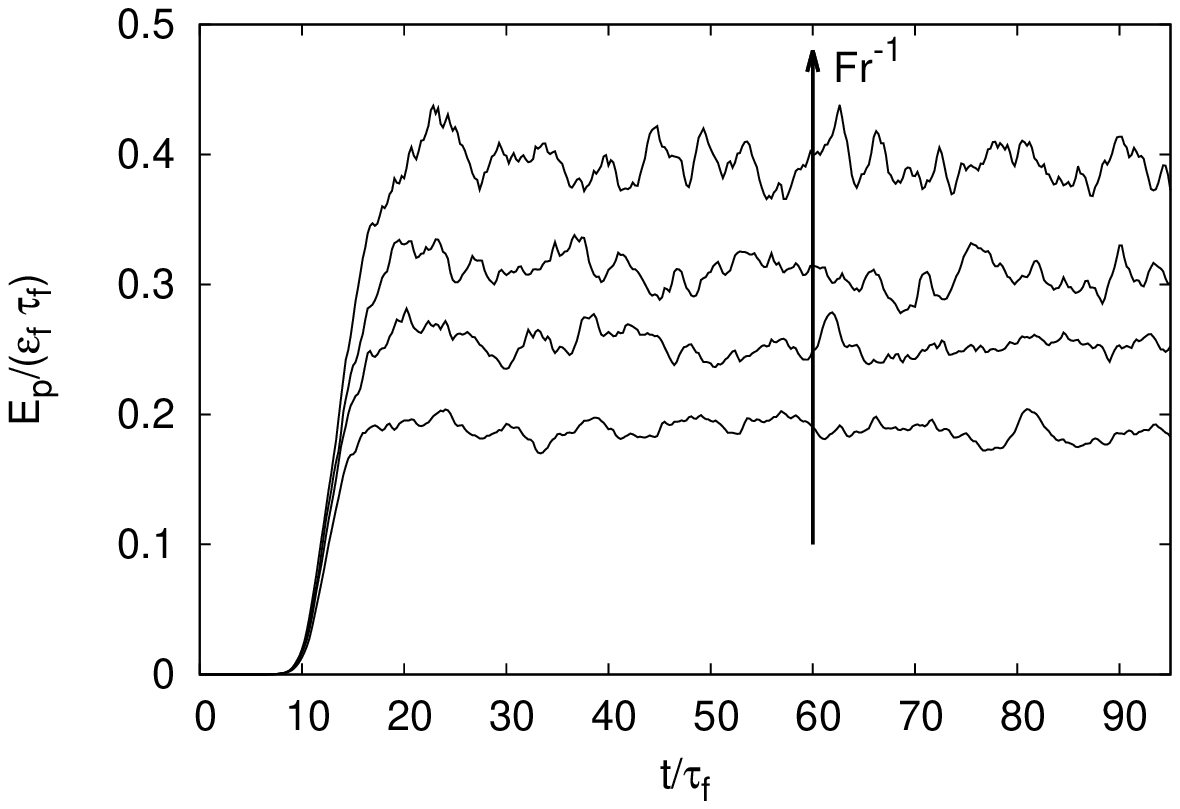}
\caption{Temporal evolution of kinetic energy $E_K(t)$ (upper) and 
potential energy $E_P(t)$ (lower) at $S=0.25$. Froude numbers are
$Fr=0.2$, $0.25$, $0.3$, $0.4$ and $Fr=\infty$.}
\label{fig1}
\end{center}
\end{figure}

For each couple of parameters $(Fr,S)$ we performed a numerical simulation 
starting from initial conditions ${\bf u}({\bf x},t=0)=0$ and 
$\phi({\bf x},t=0)=0$, 
that is with the fluid at rest and without density fluctuations. 
In the first stage ($t \leq 10 \tau_{f}$) the flow is not yet turbulent:
small scale energy dissipation is zero and kinetic energy grows at the 
the input rate $\varepsilon_{f}$. 
After this transient stage, turbulence develops and 
the potential energy $E_P$ attains statistically steady values
which increases as $Fr$ is reduced (see Figure~\ref{fig1}). 
    
The temporal evolution of the kinetic energy $E_K(t)$ is also shown in 
Figure~\ref{fig1} for different values of $Fr$.
In the turbulent stage, and in the absence of stratification, the energy 
grows linearly in time, with a rate 
$dE_K/dt = \varepsilon_\alpha<\varepsilon_{f}$ which 
is equal to the flux of the inverse energy cascade. 
By increasing the stratification this growth rate reduces, 
and eventually vanishes (for $Fr \simeq 0.2$ in this particular case).
It is worth to notice that the 
kinetic energy associated to the vertical motions 
$\langle u_3^2 \rangle$ becomes statistically constant for any $Fr$,
therefore the observed growth of kinetic energy 
is due solely to horizontal flows. 

In Table~\ref{table1} we report for each simulation the 
values of the energy growth rates $\varepsilon_\alpha$,
measured from the growth of $E_K(t)$, 
normalized with the energy input $\varepsilon_f$. 
We remind that this ratio is equal to the fraction of energy which 
is transferred to large scales producing the inverse energy cascade. 

For fixed values of $Fr$ the ratio $\varepsilon_\alpha / \varepsilon_f$ 
is a decreasing function of the aspect ratio $S$ 
(see Figure~{\ref{fig2}), 
and vanishes for a critical aspect ratio $S_c(Fr)$ which becomes smaller 
as the stratification increases. 
The behavior of $S_c$ (determined by linear interpolation of the 
lines displayed in Fig.~\ref{fig2}) as a function of $Fr$ is shown in 
Figure~\ref{fig3}. 
In the limit of vanishing stratification $Fr \to \infty$ the  
critical aspect ratio is bounded by the value $S_c \simeq 1/2$ which 
is observed in absence of stratification \citep{CMV10}. 
For sufficiently strong stratification we find that the $S_c$ 
becomes smaller and follows, for small $Fr$,
approximately the scaling $S_c \simeq Fr$. 
 
\begin{figure}
\begin{center}
\includegraphics[width=8cm]{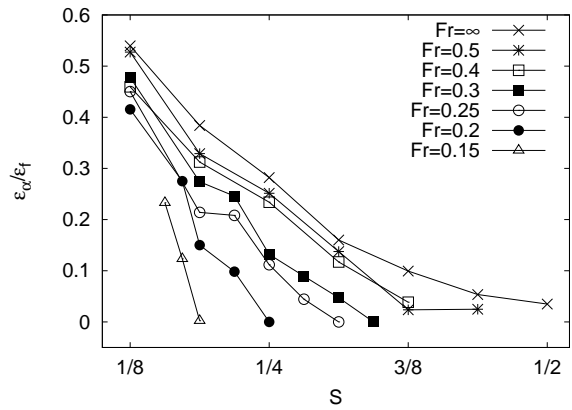}
\caption{Growth rates of kinetic energy $\varepsilon_\alpha$ 
(measured from the growth rate of $E_K(t)$)
normalized with the energy input $\varepsilon_f$, 
as a function of the aspect ratio $S=L_z/L_f$ for 
different $Fr$.}
\label{fig2}
\end{center}
\end{figure}

\begin{figure}
\begin{center}
\includegraphics[width=8cm]{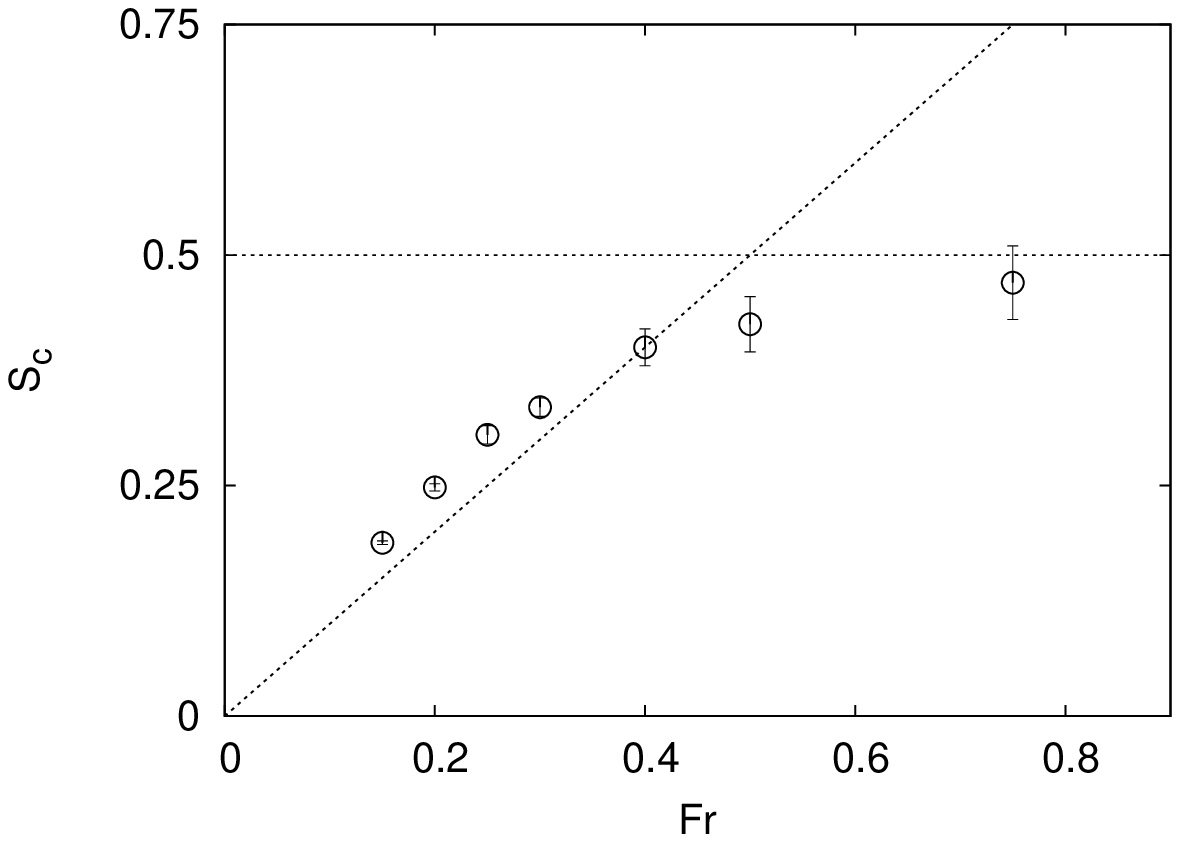}
\caption{Critical aspect ratio $S_c$, estimated from the data
in Fig.~\ref{fig2}, as a function of $Fr$.}
\label{fig3}
\end{center}
\end{figure}

The scaling of the critical aspect ratio $S_c \simeq Fr$ 
provides a crucial indication
to understand the mechanism which causes the suppression 
of the inverse energy cascade induced by stratification. 
In analogy with the aspect ratio $S$, which expresses the ratio 
between the confining scale $L_z$ and the forcing scale $L_f$, 
also the Froude number can be rewritten as the ratio between 
the characteristic vertical scale of the layered structures 
which characterize the stratified flows 
$L_v = \varepsilon_f^{1/3} L_f^{1/3}/N$ 
and the forcing scale $L_f$: 
\begin{equation}
\label{eq:ratio_fr}
\frac{L_v}{L_f} = \frac{\varepsilon_f^{1/3}L_f^{-2/3}}{N} \simeq Fr
\end{equation}
The condition to achieve a complete suppression of the inverse cascade 
$S \simeq Fr$ is therefore equivalent to the condition $L_z \simeq L_v$.  
This suggest a relation between the formation of layered 
structures and the suppression of the inverse cascade. 
In a layer with a given height $L_z$ the inverse cascade disappears 
when the stratification is sufficiently strong  
such that the typical thickness of the pancake structures 
becomes small enough to fit in the fluid layer. 
From a dynamical point of view, the suppression of the inverse cascade means 
that the kinetic energy is not transported to large scale 
and this requires that a different term in the energy transfer 
becomes relevant.

Let us therefore consider the fraction of energy which 
is transported toward small scales. 
In Figure~\ref{fig4} we show the dissipation 
rate of kinetic energy $\varepsilon_\nu$ 
and potential energy $\varepsilon_\kappa$
due respectively to the viscosity and molecular diffusivity. 
Both quantities are normalized with the total dissipation 
$\varepsilon_T = \varepsilon_\nu + \varepsilon_\kappa$. 
In the limit of vanishing stratification $Fr \to \infty$ 
one has trivially 
$\varepsilon_\nu / \varepsilon_T \to 1$ 
and 
$\varepsilon_\kappa / \varepsilon_T \to 0$. 
At increasing the stratification the fraction of energy dissipated 
by viscosity reduces, while the energy dissipation due to diffusivity grows. 
Our findings suggest that in the limit of strong stratification $Fr \to 0$ 
the two dissipations may become of the same order 
$\varepsilon_\nu \simeq \varepsilon_\kappa$. 
It is interesting to note that the ratios 
$\varepsilon_\nu / \varepsilon_T$ 
and $\varepsilon_\kappa / \varepsilon_T$ 
do not show a strong dependence on the aspect ratio $S$. 

We remark that, because from energy balance we have 
$\varepsilon_{f}=\varepsilon_{\alpha}+\varepsilon_{T}$,
the decrease of $\varepsilon_{\alpha}$ with $Fr^{-1}$ shown in Fig.~\ref{fig1}
corresponds to an increase of total dissipation $\varepsilon_{T}$
with stratification.
\begin{figure}
\begin{center}
\includegraphics[width=8cm]{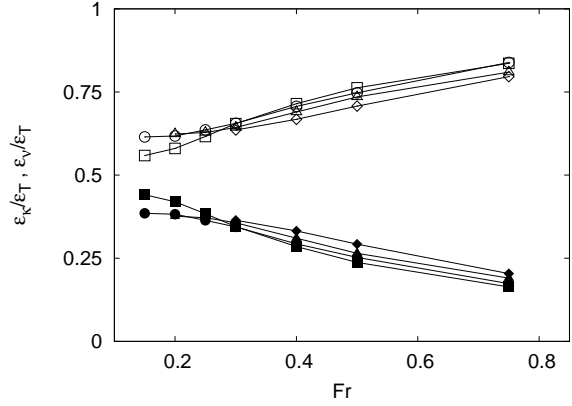}
\caption{Kinetic $\varepsilon_\nu$ (open symbols) and potential 
$\varepsilon_\kappa$ (filled symbols) energy dissipation rates 
normalized by the total dissipation
$\varepsilon_T = \varepsilon_\nu+\varepsilon_\kappa$
as a function of $Fr$ for different $S=0.125$ (squares),
$S=0.188$ (circles), $S=0.25$ (triangles) and $S=0.313$ (diamonds).}
\label{fig4}
\end{center}
\end{figure}

The exchange rate from kinetic to potential energy 
$\varepsilon_x = N \langle u_3 \phi \rangle$ is shown 
in Figure~\ref{fig5}. 
The sign of $\varepsilon_x$ is always positive, 
indicating that there is an irreversible 
conversion of kinetic energy into potential energy 
which is dissipated by the turbulent diffusivity. 
We find that the exchange rate decays approximatively 
as $\varepsilon_x \sim Fr^{-1}$, 
and grows as the aspect ratio $S$ is increased.  

\begin{figure}
\begin{center}
\includegraphics[width=8cm]{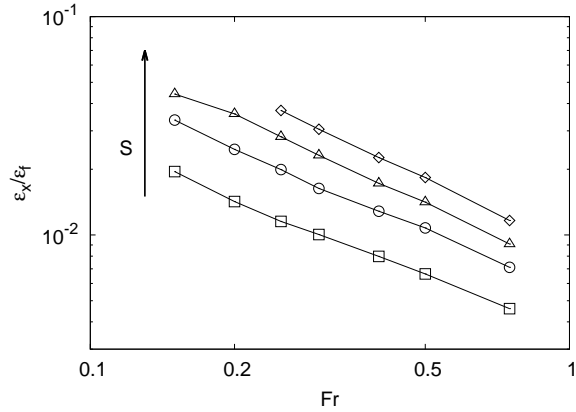}
\caption{Exchange rates between kinetic and potential energy 
$\varepsilon_x$ normalized with the energy input $\varepsilon_f$, 
as a function of $Fr$ for different $S=0.125$ (squares),
$S=0.188$ (circles), $S=0.25$ (triangles) and $S=0.313$ (diamonds).}
\label{fig5}
\end{center}
\end{figure}

\begin{figure}
\begin{center}
\includegraphics[width=8cm]{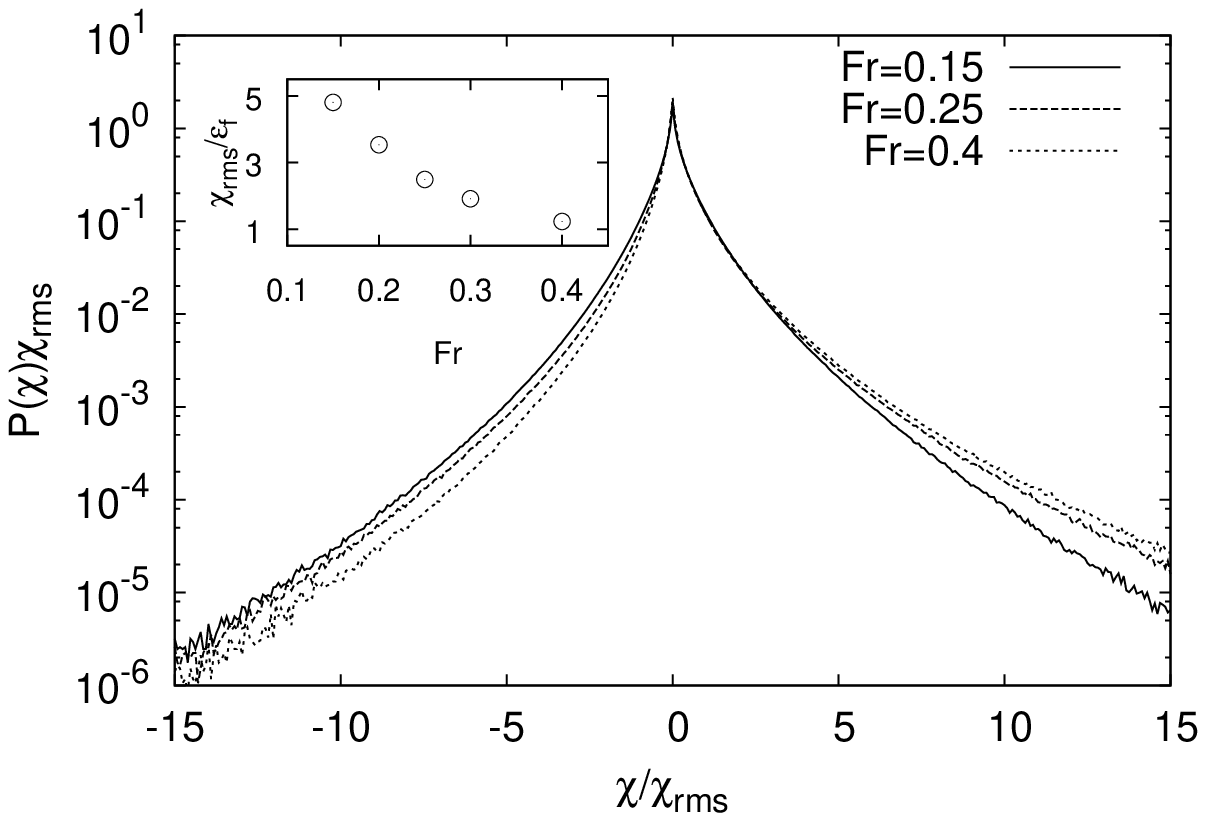}
\includegraphics[width=8cm]{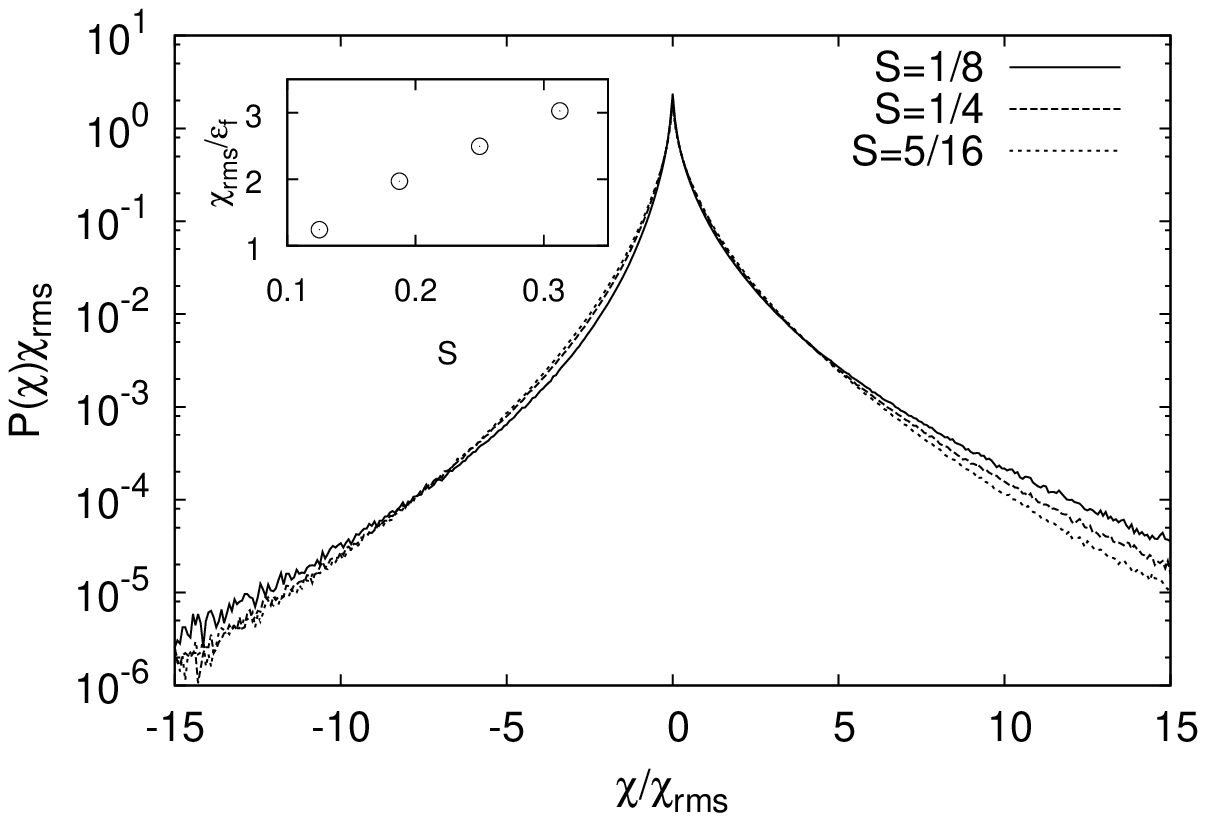}
\caption{Probability density functions of the local exchange term
$\chi$ (rescaled with the rms value) for different values of 
$Fr$ at $S=0.25$ (upper panel) and 
different $S$ at $Fr=0.25$ (lower panel). 
In the insets the dependence of the rms value $\chi_{rms}$ on
$Fr$ and $S$ is shown.}
\label{fig6}
\end{center}
\end{figure}

In Figure~\ref{fig6} we show the probability density function (PDF) 
of the local exchange rate $\chi \equiv N u_3 \phi$ (with 
$\varepsilon_x=\langle \chi \rangle$) for different values of 
$Fr$ and $S$. 
The PDFs are non-Gaussian and they are characterized by 
broad negative and positive tails. 
This implies that the mean positive value of $\chi$, 
which indicates the mean preferential 
transfer of energy from kinetic to potential,
is the results of strong cancellations between 
local events of intense energy transfer in both directions. 
We find that the asymmetry of the PDFs reduces as $Fr$ decreases, 
and that the right tail grows as the aspect ratio $S$ is reduced.
The intermittent behavior of the energy transfer 
revealed by the broad tails of the PDFs of the local exchange rate 
$\chi \equiv N u_3 \phi$
is in agreement with recent findings by \cite{RMP14} 
which have shown that also the PDFs of $\phi$ and $u_3$ 
have non-Gaussian behavior. 

\subsection{Spectral fluxes}
 
More detailed information on the mechanism of energy transfer
are provided by the inspection of the fluxes of kinetic and potential energy

\begin{eqnarray}
\Pi_{K}(k)=\int_{|\boldsymbol{q}|\leq{k}}d\boldsymbol{q}(\boldsymbol{u}
\cdot\nabla\boldsymbol{u})(\boldsymbol{q})\boldsymbol{u}^{\ast}(\boldsymbol{q})
\\
\Pi_{P}(k)=\int_{|\boldsymbol{q}|\leq{k}}d\boldsymbol{q}(\boldsymbol{u}
\cdot\nabla\boldsymbol{\phi})(\boldsymbol{q})\boldsymbol{\phi}^{\ast}(\boldsymbol{q})
\end{eqnarray}

In absence of stratification the spectral flux of kinetic energy $\Pi_K(k)$
displays two plateau (see Figure~\ref{fig7}).
At small wavenumbers $k < k_f$ the energy flux is negative, 
signaling the presence of an inverse energy cascade.  
At large wavenumbers $k > k_f$ the positive plateau of the flux 
indicates the direct energy cascade towards small scales. 

The presence of a stable stratification of density affects the double cascade 
of kinetic energy in both ranges of wavenumbers. 
At small wavenumbers the flux of the inverse cascade is reduced 
with the stratification, as $Fr$ decreases. 
Simultaneously, in the range of wavenumbers $k_f < k < k_z$ 
the flux of kinetic energy toward small scales is enhanced. 

\begin{figure}
\begin{center}
\includegraphics[width=8cm]{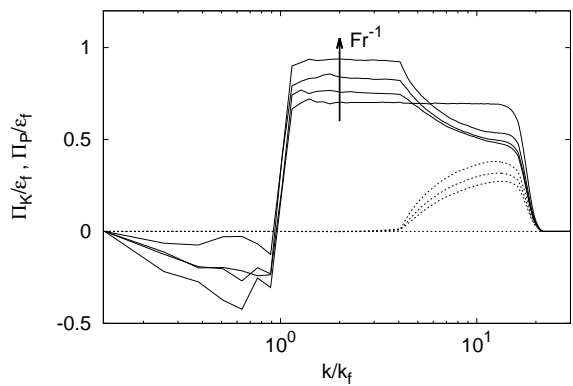}
\caption{Spectral fluxes of kinetic energy (solid lines) and potential 
energy (dotted lines) for different values of Froude number 
($Fr=\infty$, $Fr=0.3$, $Fr=0.25$, $Fr=0.2$ from bottom to top) 
and fixed aspect ratio ($S=1/4$)}
\label{fig7}
\end{center}
\end{figure}

At large wavenumbers $k > k_z$ 
(larger than the wavenumber associated 
to the thickness of the fluid layer $k_z = 2\pi /L_z=k_f/S$) 
we observe the development of a cascade of potential energy, 
whose flux grows as $Fr$ decreases. 
The direct cascade of potential energy develops 
at the expenses of the kinetic energy, whose flux shows 
a significative reduction at large wavenumbers $k > k_z$. 
Remarkably, at very large wavenumbers $k \gg k_z$,  
the fluxes of kinetic energy are always smaller 
than in the non-stratified case, although they display a weak growth 
at increasing stratification. 
This indicates that the main effect of the 
vertical shears associated to the layered structures in stratified flows
is to promote the conversion of kinetic energy into potential energy,
rather than to cause an enhancement of the viscous dissipation. 

The conversion of kinetic energy into potential energy is a process 
which requires to move a parcel of fluid in the vertical direction. 
Its efficiency is therefore 
strongly affected by the thickness of the fluid layer. 
The space required in the vertical direction to achieve the 
maximum conversion is of the order of the thickness 
of the pancake structures $L_v$. 
If the depth of the fluid layer $L_z$ is thinner than $L_v$ 
it is possible to achieve only a partial conversion, 
allowing for the survival of a remnant inverse energy cascade 
toward large scales. 
The inverse cascade is completely suppressed when the layer is 
sufficiently thick 
or the stratification is sufficiently strong, such that $L_z > L_v$, 
allowing for the maximum conversion of kinetic energy into potential energy.  
This picture is consistent with the scaling $S_c \simeq Fr$ 
observed in Figure~\ref{fig3}. 

In absence of stratification, a crucial role in determining the ratio between 
the fluxes of kinetic energy of the direct and inverse cascade 
is played by the phenomenon of vortex stretching. 
In three-dimensional flows, the vortex stretching is responsible 
for the production of enstrophy which is related to the rate of 
viscous dissipation of energy and therefore to the flux of energy in the direct cascade. 
Conversely, in ideal two-dimensional flows, the term responsible 
for the vortex stretching vanishes, and enstrophy becomes an inviscid 
invariant. 
The joint conservation of enstrophy and energy causes 
the reversal of the energy cascade which is transferred toward large scales. 

In the case of a fluid layer with a finite thickness $L_z$ smaller
than the forcing scale $L_f$ we observe the phenomenon shown in
Figure~\ref{fig8}. The spectral flux of enstrophy $\Pi_Z$ and 
the spectral production of enstrophy $\Sigma_Z$ are defined as
\begin{eqnarray}
\Pi_{Z}=\int_{|\boldsymbol{q}|\leq{k}}d\boldsymbol{q}(\boldsymbol{v}
\cdot\nabla\boldsymbol{\omega})(\boldsymbol{q})\boldsymbol{\omega}^{\ast}(\boldsymbol{q})
\\
\Sigma_{Z}=\int_{|\boldsymbol{q}|\leq{k}}d\boldsymbol{q}(\boldsymbol{\omega}
\cdot\nabla\boldsymbol{v})(\boldsymbol{q})\boldsymbol{\omega}^{\ast}(\boldsymbol{q})
\end{eqnarray}

In the non-stratified case ($Fr = \infty$) 
the enstrophy flux is constant in the range $k_f < k < k_z$, 
indicating the presence of a direct cascade of enstrophy,
consistent with a quasi-two-dimensional phenomenology.
The enstrophy production is activated only at $k > k_z$ 
where therefore the enstrophy flux is not conserved.
In the stratified case ($Fr =0.2$ in Figure~\ref{fig8}) 
the enstrophy flux remains unchanged 
in the range $k_f < k < k_z=k_f/S$. At higher wavenumber 
we do not observe a relevant increase of the production of enstrophy, 
which remains similar to the non-stratified case. 
On the contrary, the flux of enstrophy is reduced, 
signaling that a fraction of the enstrophy generated 
by the vortex stretching is spent in the process of 
conversion of kinetic energy into potential energy.

\begin{figure}
\begin{center}
\includegraphics[width=8cm]{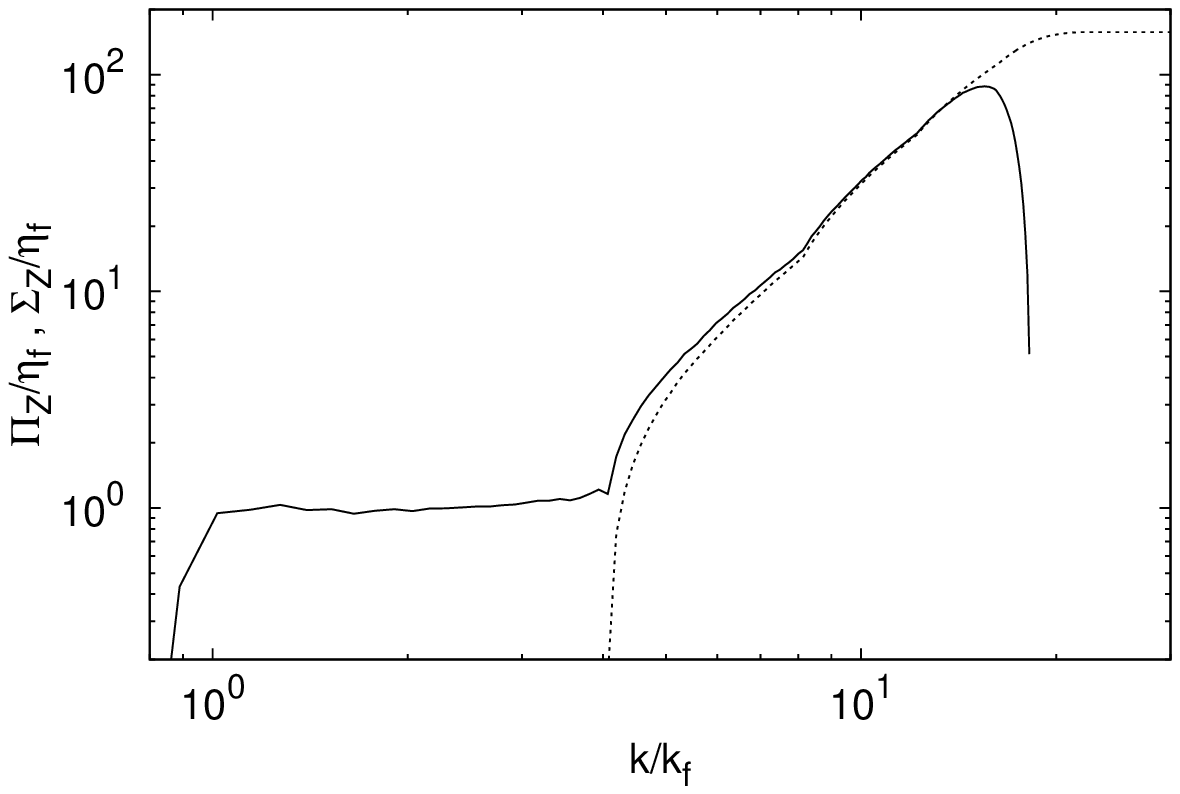}
\includegraphics[width=8cm]{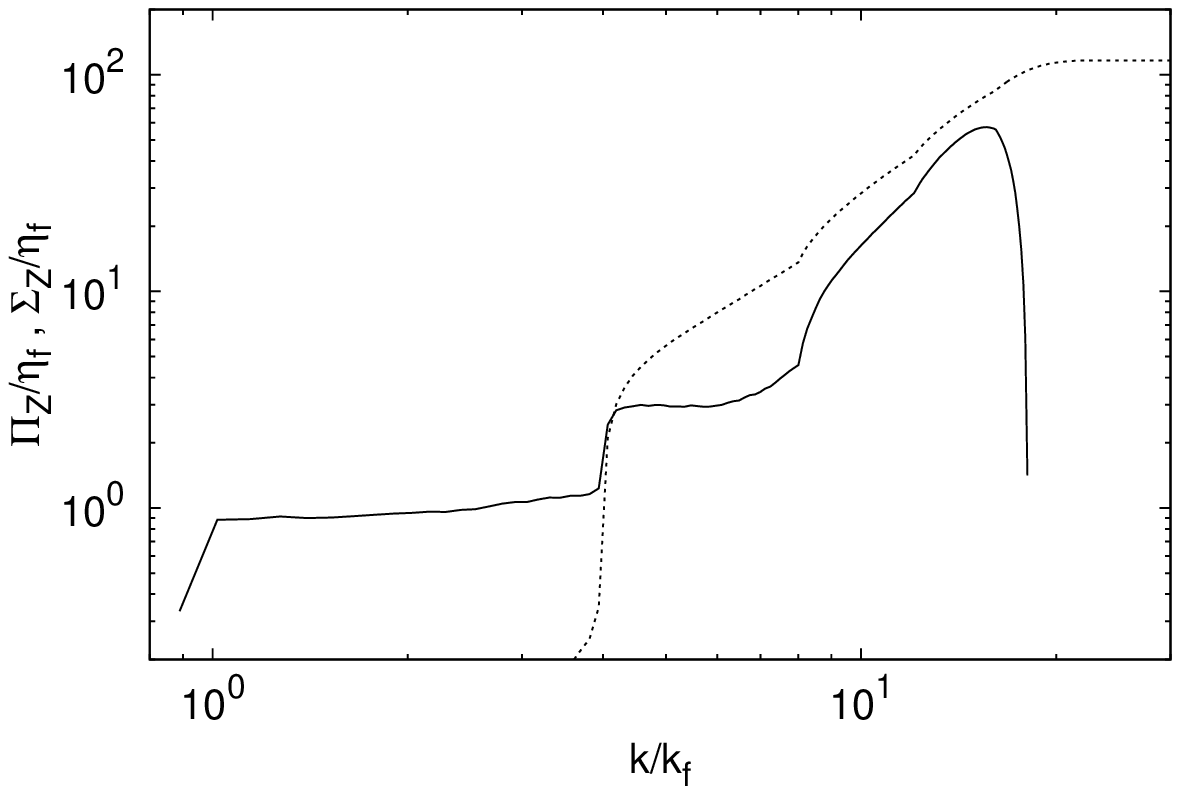}
\caption{Spectral flux of enstrophy \(\Pi_{\omega}\) (solid line) and
vortex-stretching \(\Sigma_{\omega}\) (dashed line) for simulations
at $S=0.25$ and $Fr=\infty$ (upper panel) and $Fr=0.2$ (lower panel).
All quantities are normalized by the
enstrophy input by forcing.}
\label{fig8}
\end{center}
\end{figure}

\section{Conclusions}
\label{sec4}

We have investigated the direction of the energy flux 
in a set of numerical simulations of a thin layer of stratified fluid
for different values of stratification (Froude number $Fr$) and 
thickness (aspect ratio $S$). 
We have shown that, in general, stratification reduces the intensity
of the inverse cascade and, consequently, the critical value of
stratification $S_c$ at which the inverse flux vanishes.

For small values of $Fr$, this critical number is found to grow
approximatively as $S_c \simeq Fr$. This fact supports the picture
by which the inverse cascade vanishes when the thickness of the 
pancake structures in the flow is sufficiently small to fit
into the fluid layer. For larger values of $Fr$, the critical
aspect ratio is found to recover the unstratified limit.

A spectral analysis of the kinetic and potential energy fluxes 
shows that the suppression of the inverse cascade of kinetic energy
is accompanied by the generation of a direct cascade of potential energy which
becomes an alternative channel for the transfer and dissipation
of injected energy.

Our findings open new pespectives for both experimental
and theoretical research in turbulence.

Stable stratification of density has been often used in   
experimental setup of electromagnetically-forced thin fluid layers
with the purpose of suppressing vertical motions and enhancing   
the two-dimensionality of the flow.
It would be interesting to observe experimentally
the suppression of the inverse cascade at increasing the stratification,
and to compare the results with our numerical findings.

From a theoretical point of view it would be extremely useful to have
a predictive phenomenological model which allows to determine the fraction
of energy which is drained by the direct cascade of potential energy,
and therefore to have a quantitative prediction for the suppression
of the inverse energy cascade as function of both the aspect ratio
and the stratification.    


\begin{acknowledgements}
Authors are grateful to Filippo De Lillo and Enrico Deusebio for fruitful
discussions. Numerical simulations were performed on the INFN Turbofarm cluster
in Turin, Italy.\\
\end{acknowledgements}

\appendix
\section{K\'arm\'an-Howart-Monin equation for stratified turbulence}
\label{appA}
In this Appendix we derive a set of generalized K\'arm\'an-Howart-Monin (KHM) equation 
for stably stratified turbulence. 

We define the velocity and scalar two-point correlation functions as:
\begin{equation}
C_{2}^{(u)}({\bf x},t) \equiv \langle {\bf u}({\bf x},t) \cdot 
{\bf u}({\bf 0},t) \rangle
\label{eq:A.1}
\end{equation}
\begin{equation}
C_{2}^{(\phi)}({\bf x},t) \equiv \langle \phi({\bf x},t) 
\phi({\bf 0},t) \rangle
\label{eq:A.2}
\end{equation}
and the correlation among the two fields, representing the exchange of
energy between kinetic and potential
\begin{equation}
C_{1,1}^{(u_3,\phi)}({\bf x},t) \equiv \langle u_3({\bf x},t) 
\phi({\bf 0},t) \rangle \, .
\label{eq:A.3}
\end{equation}
Dissipation of kinetic and potential energy are defined as
($\alpha,\beta=1,2,3$ and we sum over repeated index)
\begin{equation}
D^{(u)}({\bf x},t) \equiv 
2 \nu \langle (\partial_{\alpha} u_{\beta})({\bf x},t) 
(\partial_{\alpha} u_{\beta})({\bf 0},t) \rangle
\label{eq:A.4}
\end{equation}
\begin{equation}
D^{(\phi)}({\bf x},t) \equiv 
2 \kappa \langle (\partial_{\alpha} \phi)({\bf x},t) 
(\partial_{\alpha} \phi)({\bf 0},t) \rangle
\label{eq:A.5}
\end{equation}
and cross dissipation as
\begin{equation}
D^{(u,\phi)}({\bf x},t) \equiv 
2 \nu \langle (\partial_{\alpha} \phi)({\bf x},t) 
(\partial_{\alpha} u_3)({\bf 0},t) \rangle +
2 \kappa \langle (\partial_{\alpha} u_{3})({\bf x},t) 
(\partial_{\alpha} \phi)({\bf 0},t) \rangle 
\label{eq:A.6}
\end{equation}

The correlation of the two-dimensional, 
two-component forcing ${\bf f}({\bf x},t)$ is given by
\begin{equation}
\langle f_{i}({\bf x},t) f_{j}({\bf 0},0) \rangle =
\delta_{ij} \delta(t) F(r_h/L_f)
\label{eq:A.7}
\end{equation}
where $i,j=1,2$ and $r_h^2=x_1^2+x_2^2$ is the separation on the horizontal
plane and $F(x)$ is the spatial correlation.

The K\'arm\'an-Howart-Monin equations will involve also third-order
structure functions. In particular we define
\begin{equation}
{\bf S}_{3}^{(u)}({\bf x},t) \equiv 
\langle [{\bf u}({\bf x},t)-{\bf u}({\bf 0},t)]
|{\bf u}({\bf x},t)-{\bf u}({\bf 0},t)|^2 \rangle
\label{eq:A.8}
\end{equation}
\begin{equation}
{\bf S}_{1,2}^{(u,\phi)}({\bf x},t) \equiv 
\langle [{\bf u}({\bf x},t)-{\bf u}({\bf 0},t)]
[\phi({\bf x},t)-\phi({\bf 0},t)]^2 \rangle
\label{eq:A.9}
\end{equation}
\begin{equation}
{\bf S}_{2,1}^{(u_3,\phi)}({\bf x},t) \equiv 
\langle [{\bf u}({\bf x},t)-{\bf u}({\bf 0},t)]
[u_3({\bf x},t)+\phi({\bf x},t)-u_3({\bf 0},t)-\phi({\bf 0},t)]^2 \rangle
\label{eq:A.10}
\end{equation}

Starting from (\ref{eq:2.1}-\ref{eq:2.2}), by exploiting homogeneity and
incompressibility, we derive a set of generalized KHM
equations, the first for kinetic energy
\begin{equation}
{\partial C_{2}^{(u)}({\bf x},t) \over \partial t} +
D^{(u)}({\bf x},t)+N C_{1,1}^{(u_3,\phi)}({\bf x},t) = 
F({\bf x}) + {1 \over 2} \bnabla \cdot {\bf S}_{3}^{(u)}({\bf x},t)
\label{eq:A.11}
\end{equation}
and the second for potential energy
\begin{equation}
{\partial C_{2}^{(\phi)}({\bf x},t) \over \partial t} +
D^{(\phi)}({\bf x},t)-N C_{1,1}^{(u_3,\phi)}({\bf x},t) = 
{1 \over 2} \bnabla \cdot {\bf S}_{1,2}^{(u,\phi)}({\bf x},t)
\label{eq:A.12}
\end{equation}
At variance with the usual Navier-Stokes equation \citep{F95},
(\ref{eq:A.11}) involves the additional term $C_{1,1}^{(u_3,\phi)}$
which represents the transfer of energy from kinetic to potential term. 
The equivalent of the KHM equation for the exchange energy reads
\begin{equation}
{\partial C_{1,1}^{(u_3,\phi)}({\bf x},t) \over \partial t} +
D^{(u_3,\phi)}({\bf x},t)-2 N C_{2}^{(u_3)}({\bf x},t) +
2 N C_{2}^{(\phi)}({\bf x},t) = 
{1 \over 2} \bnabla \cdot {\bf S}_{2,1}^{(u,\phi)}({\bf x},t)
\label{eq:2.13}
\end{equation}

Adding up (\ref{eq:A.11}) and (\ref{eq:A.12}), this energy exchange terms
cancels and we get the KHM relation for the total energy
\begin{eqnarray}
&& {\partial \over \partial t}\left(C_{2}^{(u)}({\bf x},t)+
C_{2}^{(\phi)}({\bf x},t \right) +
\left(D^{(u)}({\bf x},t)+D^{(\phi)}({\bf x},t) \right)= \nonumber \\
&& 
F({\bf x}) + {1 \over 2} \bnabla \cdot \left(
{\bf S}_{3}^{(u)}({\bf x},t)+{\bf S}_{1,2}^{(u,\phi)}({\bf x},t) \right)
\label{eq:A.13}
\end{eqnarray}
For ${\bf x} \to 0$, at finite $\nu$ and $\kappa$,
(\ref{eq:A.13}) gives the energy balance equation
\begin{equation}
{d \over dt} (E_k + E_p) + \varepsilon_{\nu} + \varepsilon_{\kappa} = 
\varepsilon_f
\label{eq:A.14}
\end{equation}
where $\varepsilon_f=F(0)/2$ is the energy input and $\varepsilon_{\nu}$ 
and $\varepsilon_{\kappa}$ are the viscous and diffusive energy
dissipations respectively.
In the absence of stratification, usual Navier-Stokes equations in 3D
reach a steady state in which $\varepsilon_{\nu}=\varepsilon_f$ and
$dE/dt=0$. This is not the case in 2D, where the inverse cascade
transfers kinetic energy to large scales where viscous dissipation
is not effective (in the limit of large Reynolds numbers). 
Previous investigations and our numerical simulations
show that also in presence of stratification total energy reaches
a steady state, indicating the absence of inverse cascade. Therefore
in the following we will assume stationarity.

Starting from equation~(\ref{eq:A.11}) and
assuming stationarity and dissipative anomaly for the kinetic energy
in the limit of vanishing viscosity, one gets:
\begin{equation}
\lim_{\nu \to 0} \lim_{|{\bf x}| \to 0}
\left\{
D^{(u)}({\bf x})+N C_{1,1}^{(u_3,\phi)}({\bf x}) - F({\bf x})
\right\}
=0
\label{eq:A.15}
\end{equation}
which means that the energy input is partly dissipated and partly
transferred to the potential energy:
\begin{equation}
D^{(u)}({\bf 0})+N C_{1,1}^{(u_3,\phi)}({\bf 0}) = F({\bf 0})
\label{eq:A.16}
\end{equation}
Furthermore, taking the same limits of equation~(\ref{eq:A.11})
but now in inverse order yields
\begin{equation}
{1 \over 2}\lim_{|{\bf x}| \to 0}\lim_{\nu \to 0} 
 \bnabla \cdot {\bf S}_{3}^{(u)}({\bf x},t)= -\left\{F({\bf 0}) - N C_{1,1}^{(u_3,\phi)}({\bf 0})\right\}
\,\leq\,0
\label{eq:A.16.bis}
\end{equation}
The negative value of the divergence is the hallmark of the direct energy cascade. 
The relation is thus agreement with the presence exhibited in Figure~(\ref{fig7}) 
of a direct energy cascade for any value of $Fr$.

In a similar way, assuming stationarity and
dissipative anomaly for the potential energy
in the limit of vanishing diffusivity in equation~\ref{eq:A.12}
one gets:
\begin{equation}
\lim_{\kappa \to 0} \lim_{|{\bf x}| \to 0}
\left\{
D^{(\phi)}({\bf x})-N C_{1,1}^{(u_3,\phi)}({\bf x})
\right\}
= 0
\label{eq:A.17}
\end{equation}
and hence:
\begin{equation}
D^{(\phi)}({\bf 0)})= N C_{1,1}^{(u_3,\phi)}({\bf 0)}) \ge 0
\label{eq:A.18}
\end{equation}
This means that the cross-correlation $C_{1,1}^{(u_3,\phi)}$,
being equal to the dissipation of potential energy, is positive definite.
Therefore, it acts as a source term for the potential energy
and as a dissipation for the kinetic energy.


\bibliographystyle{jfm}

\end{document}